# An Optimizing Just-In-Time Compiler for Rotor

João H. Trindade, José C. Silva

ISEL – DEETC
R. Conselheiro Emídio Navarro, 1
1950-062 Lisboa, Portugal
`jtrindade.net@iol.pt, jsilva.net@iol.pt`

**Abstract.** The Shared Source CLI (SSCLI), also known as Rotor, is an implementation of the CLI released by Microsoft in source code. Rotor includes a single pass just-in-time compiler that generates non-optimized code for Intel IA-32 and IBM PowerPC processors. We extend Rotor with an optimizing just-in-time compiler for IA-32. This compiler has three passes: control flow graph generation, data dependence graph generation and final code generation. Dominance relations in the control flow graph are used to detect natural loops. A number of optimizations are performed during the generation of the data dependence graph. During native code generation, the rich address modes of IA-32 are used for instruction folding, reducing code size and usage of register names. Despite the overhead of three passes and optimizations, this compiler is only 1.4 to 1.9 times slower than the original SSCLI compiler and generates code that runs 6.4 to 10 times faster.

## Introduction

Rotor's JIT compiler (FJIT) translates CIL [1] to native code with a single pass over the code [2]. This is done in a loop, reading CIL instructions and using them to select an entry in a switch. Each entry emits the block of native instructions equivalent to its CIL instruction. As a consequence, emitted code will have an execution model equivalent to the CIL stack model: intermediate results are stored in the stack for later use, and each instruction reads its parameters from the stack; if some result is used more than once, it is kept in a local variable.

This way of structuring a compiler has two major advantages and one major drawback. On one hand, it makes it easier to build a compiler, which is both highly portable and fast. On the other hand, code generated by this compiler will perform poorly.

By sacrificing portability, it is possible to build an optimizing compiler that generates better code by taking full advantage of the instruction set and other characteristics of the target machine. Unfortunately, this will not only sacrifice portability, but also compilation speed.

We describe a new compiler for Rotor (AJIT) that tries to keep a balance between compilation speed and generated code quality. We follow the principles described by Chen and Olukotun for microJIT [3], a Java just-in-time compiler. The basic idea is to keep the number of passes low and to integrate a selected group of optimizations into these passes.

## Compiler Structure

AJIT does three major passes over the code. The first performs control flow analysis, dividing the code in basic blocks and detecting loops. The second builds a data dependence graph for each basic block, thereby creating an intermediate representation for the code based on directed acyclic graphs (DAGs). This phase includes most of the optimizations performed by the compiler. Finally, the third pass transforms the intermediate representation into native code, performing register allocation and taking advantage of the rich address modes of IA-32 architecture.

## Control Flow Analyzer

Basic blocks (BBs) are defined on top of the notion of leader:

1. the very first instruction is a leader;
2. a branch target is a leader;
3. the instruction after a branch is a leader.

Now we can define a basic block as a sequence of instructions that starts in a leader and finishes in the instruction immediately before the next leader. The algorithm presented in Fig. 1 divides code into basic blocks and constructs a Control Flow Graph (CFG) with one single pass over the code.

The algorithm goes through the sequence of instructions only once. It keeps a list of identified leaders to speed up the process of detecting the end of the current block. Branching forward may create a new leader; branching backward may break an existing block in two. Whenever the offset of the current instruction is equal to the offset of the next identified leader, the current block is closed and a new one is created.

In order to detect the end of the current block, it suffices to compare the offset of each instruction with the offset of the next identified leader. The latter is kept in a local variable, allowing for efficient comparison. The auxiliary function *CreateLeader* inserts a new leader into the list of leaders, if one does not exist at that position. The auxiliary function *CreateBlock* creates a new basic block at the indicated position, if it does not already exist. It may happen that a backward branch has its target inside an existing block. In that case, the existing block is shortened and a new block is inserted between the old block and its successors. This way, it is possible to keep the old connections from the predecessors to the old block, and just transfer its list of successors to the new block. After the split, the new block is the only successor of the old block, and the new block has two predecessors: the old block and the one that caused the split with a backward branch.

The resulting CFG has only information about each block's successors. Information about predecessors is added during the loop detection phase. However, the number of predecessors is counted during CFG construction. This way, information about predecessors may be kept in arrays, instead of dynamic structures.

```
first_block := CreateBlock(first_instruction)
last_block := CreateBlock(method_end)
current_block := first_block
next_leader := last_block.leader
previous_instruction := nil
for each instruction, i, in current_block
  if next_leader == i
    current_block.end := i
    if previous_instruction == conditional_branch or switch
      current_block.successors += next_leader.block
    current_block := next_leader.block
    ++next_leader
  switch i
    case BR, LEAVE
      current_block.successors += CreateBlock(i.target)
      CreateLeader(next_instruction)
    case SWITCH
      for each case, c
        block_current.successors += CreateBlock(c.target)
      CreateLeader(next_instruction)
    case RET, THROW, RETHROW, JMP
      CreateLeader(next_instruction)
  previous_instruction := i
```

**Fig. 1.** Algorithm for building the Control Flow Graph

## Loops

After the major pass to build the CFG, another phase detects loops on it. This is important because it is highly likely that any program will spend most of its execution time around some loop or loops. Hence, heavy optimizations should concentrate on loops. At the moment, AJIT does not include loop optimizations, but the loop detector is already there.

## Dominators

In a control flow graph, a dominator D of a basic block B is a basic block that must be traversed to reach B from the first basic block R.

For calculating dominators, we use the algorithm described by Cooper et al. [4]. As the algorithm requires nodes to be numbered, we start by numbering the nodes of the CFG in depth-first order.

Once basic blocks are numbered and dominators determined we are able to detect loops.

**Natural Loops**

AJIT detects natural loops within a control flow graph. A natural loop is detected when a back edge links a basic block to one of its dominators (including itself). Non-natural loops are not detected, but they are not present in structured languages as the ones supported by the CLI, and can only be produced by direct coding in CIL.

Natural loop detection produces a loop tree. All tree nodes are loops and have as many child nodes as needed. The leaf nodes are the inner loops. For implementation convenience, the root node is a "virtual loop" representing the whole method. Each node is connected to its parent, siblings, direct children, and to its first basic block.

A natural loop is detected when a basic block dominates one of its predecessors. Fig. 2 shows the natural loop detection algorithm:

```
for each basic block, bb1, in reverse postorder:
  if bb1.num_predecessors > 0 then:
    for each bb1.predecessor, predBb1 :
      /*bb_dfn: bb's depth first number*/
      curIDom := predBb1.bb_dfn;
      while curIDom < bb1.bb_dfn :
        curIDom := IDOM[curIDom];
      if bb1.bb_dfn == curIDom then:
        /* bb1 is a dominator; we found a natural loop. */
        newLoop = new_loop();
        newLoop.header := bb1;
        getBBsWithinNaturalLoop(Root,newLoop,bb1,predBb1);
```

**Fig. 2.** Natural loop detection algorithm

We use the vectors of immediate dominators to detect whether a basic block dominates another one. In order to get all dominators for a given basic block, B, we start by indexing the vector with B and successively indexing with the result from the previous indexation. This process consists of a straightforward application of the following dominators' property:

− If D immediately dominates C then D dominates C.
− If D immediately dominates a dominator of C then D dominates C.

Another important characteristic of which we take advantage in our algorithm is that postfix numbering implies that a dominator always has a higher number than a basic block which it dominates. Taking advantage of that, we just need to repeat the successive indexing over the immediate dominators' vector while we retrieve a lower number than the hypothetical dominator's basic block number.

Once a natural loop is found, we gather all its constituent basic blocks by following all predecessors of the loop's tail (the basic block where we detected the back edge to the dominator) until we reach the dominator.

## Data Dependence Analyzer

After the flow graph analysis completion another major pass is needed to perform data dependence analysis. This pass simulates execution using a virtual stack. CIL instructions are fully parsed and transformed into Intermediate Representation (IR) instructions.

The adopted IR instruction set is based on the CIL instruction set. New instructions were added to improve some optimizations. Part of these new instructions result from splitting existing CIL instructions to expose sub-operations like null reference check or index out of bounds check in array operations. New IR instructions maintain explicit data dependence information by having pointers to other IR instructions as their arguments. Program sequence information is also maintained by linking IR instructions in a list. This information is important to guarantee exception order, as IR instructions that throw exceptions cannot be reordered. Fig. 3 shows the data structure adopted in AJIT for IR instructions.

Most optimizations are done in this pass while issuing IR instructions. This reduces the number of major passes over the code, improving the speed of just-in-time compilation.

The transformation from CIL to IR is done for each basic block in a reverse postorder traversal over all basic blocks. This order ensures that, in almost all the cases, when a basic block is processed all their predecessors have already been processed. Some optimizations that require predecessor's information are then possible.

```
struct Instruction {
  Opcode      opcode; // instr code
  Instruction[] in; // in parameters
  Type        type;   // result type
  Flags       flags;  // volatile, tail,
                      // unaligned,
                      // throws,
                      // embeddable.
  BasicBlock  bb;     // basic block
  Instruction prev;   // prev instr.
  Instruction next;   // next instr.
  int         counter; // usage counter
  MDData      mdData; // machine
                      // dependent data
};
```

**Fig. 3.** Data structure for IR instructions

CIL virtual stack is used to simulate the execution of CIL instructions and to determine the data dependencies between IR instructions. Whenever an IR instruction produces a result, a pointer to that IR instruction is placed in the stack. The next IR instruction with one or more operands will consume that stack entry. Therefore, the new IR instruction will point directly to the IR instruction that produced its operand.

The general algorithm for constructing the data dependence graph is:

1. Create an Instruction instance with data retrieved from CIL instruction.

2. Initialize usage counter with 1.
3. Append new Instruction instance to current basic block instruction list.
4. Remove from stack as many pointers to Instruction as the number of instruction input parameters, storing them into `in` instruction's field.
5. If instruction generates a result:
    5.1.  Determine result type and store it into instruction's type field
    5.2.  Add a stack entry with a pointer to new Instruction

As shown in Fig. 3, IR instructions have a usage counter. The usage counter default value is 1. This value could change due to the action of some CIL instructions like DUP or POP, which may increment it or decrement it, respectively. An instruction's usage counter is also incremented when, as part of the optimization process, new instructions are made to point at it. The counter is decremented when one of the instructions that point at it is eliminated.

**Basic Block Adaptation**

After processing a basic block, some elements may remain on stack. In this case, when evaluating the successor basic blocks, the stack must be initialized with these remaining elements. As each successor may have several predecessors, each of which terminating with the same number of items on stack, it is impossible to track a single source instruction for each of these elements.

For that reason we have intermediate representation instructions for storing and retrieving these elements from temporary locations in memory. We've called these instructions STTMP and LDTMP, respectively.

If, after processing a basic block, there are elements on stack, we emit intermediate representation STTMP instructions to store them on temporary locations allocated on the local stack frame. For all basic blocks, if their predecessors stored elements with STTMP, an equivalent number of LDTMP is emitted at the head of the block to retrieve the values from the temporaries.

STTMP instructions are emitted at the earliest safe position in the basic block. Ideally, this should be right after the source instruction, which may allow the code generator to use the rich address modes of IA-32, such that the source instruction generates its result directly into the temporary location.

**Optimizations**

AJIT looks for optimizations while it decodes CIL instructions and builds the data dependence graph. It does not yet support global analysis, thus data dependence analysis and optimizations are being done basic block by basic block.

In presence of logical or arithmetic operations with one or more constant operands, AJIT performs constant folding and algebraic simplification whenever possible If a logical or arithmetic operation with constants results in an exception (e.g. divide by zero) it gets replaced by a `throw` IR instruction for the appropriate exception.

In a sequence of CIL instructions it is common to find redundant loads and stores. Redundant loads and stores are, in normal conditions, identified as:

1. A second load local variable instruction (LDLOC) in a basic block.
2. An LDLOC instruction after a store local variable (STLOC) of the same variable.
3. Two stores over the same local variable. If LDLOC instructions between the two STLOCs can get the value of the local variable without reading it from memory, then the first STLOC becomes unnecessary.

In order to minimize the number of load and store instructions over local variables and arguments, we implemented a virtual data repository mechanism. Two separate virtual data repositories are instantiated, one for arguments and another one for local variables. These repositories have as many lines as the number of variables/arguments. Each line may be empty (pointing to null) or pointing to a load instruction (LDLOC or LDARG) or a store instruction (STLOC or STARG). Each line in the repository points to the last operation done over the corresponding variable/argument. As we only do local analysis, these repositories are re-initialized pointing to null whenever a new basic block starts.

Fig. 4 shows an example of unnecessary load and store elimination using a virtual data repository.

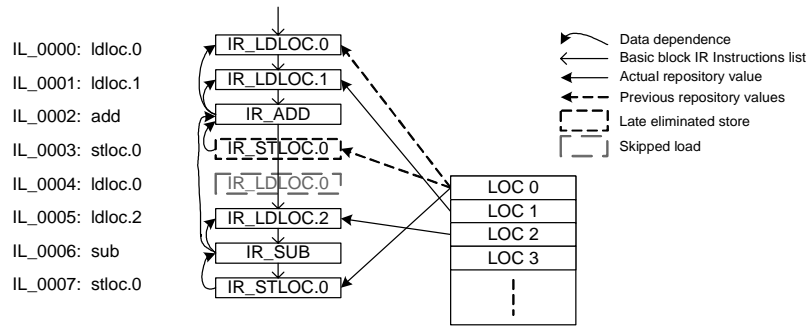

**Fig. 4.** Example of unnecessary load and store elimination using virtual data repository

We detect dead code whenever an instruction's result is not used anymore. When a usage counter is decremented due to optimizations and reaches zero, no dependent instruction remains. Dead code elimination was implemented in the following way:

1. If a dead instruction has arguments, iterate over them and decrement the usage counter of each instruction. This process originates new dead code elimination tasks whenever one or more of those usage counters reaches zero.
2. Remove the dead instruction from the basic block's list of instructions.

In this process, instructions that may throw exceptions cannot be eliminated, unless they become unreachable. Regardless of whether results are used or not, we have to guarantee exceptions' order. *Call* instructions cannot be eliminated either, even if their return values are not used, because calling a method may cause side effects. Anyway, *call* instructions are already protected by the previous case because they may throw exceptions.

Sequences of load–operation–store instructions are common in CIL due to the stack-based nature of the language. In this scenario, we want to achieve the following goals when generating native code:

1. Avoid explicit load instructions whenever the result has only one usage and we can take advantage of CISC processors' addressing modes to embed the load on the operation's native instruction. Embedding loads within operation instructions reduces code size, usage of register names, and spilling needs.
2. Explicitly issue a load instruction when its usage counter is greater than 1 in order to place the value in a register, thereby reducing variable/argument load latency.
3. Avoid explicit store instructions whenever the destination can be directly specified within the operation's instruction.

A *CanEmbed* flag was added to IR instructions to support the first goal. The code generation pass will skip load instructions marked with *CanEmbed*. When issuing an IR instruction that has a load instruction as one of its arguments, we set that load instruction's *CanEmbed* flag if its usage counter is 1 and the respective entry at the virtual repository still points at it (variable value did not change meanwhile). If the usage counter is later incremented, the *CanEmbed* flag is reset, achieving the second goal.

When issuing a store instruction, we analyze the possibility of embedding it within the instruction that generated the value to be stored, thereby achieving the third goal. A store cannot be embedded if a load or store over the same variable occurs between the instruction that produced the value and the store instruction. This is verified by iterating backward over the basic block looking for the IR instruction from which it depends. This backward iteration is limited to 8 instructions. As demonstrated in [6], for the Java virtual machine, a distance of more than 6 instructions between the operation and the store has almost null probability.

## Code Generation

The last major pass transforms IR into IA-32 native code. The code generator traverses all CFG's basic blocks, in the same order as they were in the original CIL, and each IR instruction is translated to the corresponding native code. Register allocation is performed at the same time, with the most recently released register being the next to reuse or, in case there are no free registers, releasing the least recently occupied register. When an occupied register is released for allocation, the value it contains is transferred into a memory position. This action is known as spilling.

### Stack Frame

We adopt, as a base, FJIT's stack frame, which is then extended with: two slots for ebx and edi; a zone for temporaries (for BB adaptation); and a zone for spilling.

The zone for temporaries has a fixed size, such that it is able to hold the maximum size of values left on stack by any of the method's basic blocks. This size is calculated during data dependence analysis. The zone for spilling is managed like a stack.

**Code Generation**

For each instruction, the following steps are taken:

1. Determine the IA-32 instruction type: register - register; register - immediate; memory - register; memory - immediate; register - memory.
2. Allocate registers for the instruction.
3. Emit the instructions for placing data on the allocated input registers.
4. Emit the native instructions.
5. Decrement the usage counters for the instructions associated with the input registers. Those reaching zero allow the corresponding registers to be set free.
6. If the instruction has an embeddable store, but it was impossible to embed it, the store instruction is emitted separately.

For each emitted branch, we record the address of the physical instruction and the target basic block number. The start address of each basic block is also recorded. In the end, an additional pass over the generated code corrects all branch offsets.

**Register Allocation**

Register allocation assigns registers to the arcs of the data dependence graph. Due to some IA-32 instructions that require specific registers (e.g. idiv, or the shift instructions), it is not always possible to assign the same register to both ends of an arc. In that case, the arc must be broken in two, and an additional node is inserted for moving data between the required registers. As for instructions that operate directly in memory, the arcs corresponding to memory accesses do not have registers assigned.

In AJIT, register allocation does not have a phase of its own. It occurs simultaneously with code generation, using simple criteria for assigning registers to arcs: if there are registers available, the one that has been released most recently is assigned; otherwise, we choose the least recently occupied register. We assign the most recently released register to minimize the total number of registers used in one single method. By assigning the one that has been occupied for the longest time, we are using the following heuristic: if it hasn't been used so far, it won't be used for some time more.

When allocating registers, we take several factors into account:

- in general, in IA-32, input register #1 is the same as the exit register;
- input register #1 must be different from input register #2 (unless they have the same source);
- some of the registers may be pre-allocated;
- the input register of an instruction that uses the result of the current instruction may be pre-allocated;
- one, or both, of the entries may be an embeddable load;

- the instruction may be connected to an embeddable store;
- one of the entries may be a constant;
- one of the entries may be spilled;
- the instruction associated to input #1 may have its usage counter with a value greater than 1, in which case its value must be saved into another register, because the current one will be eliminated by the execution of the current instruction.

Given the number of factors that must be considered during register allocation, this task has been divided in four steps:

1. Identify the type of IA-32 instruction.
2. Propagate pre-allocation information.
3. Allocate registers
4. Deal with additional data movement.

In step 1, we start by detecting whether an immediate value may be used on input #2. This is true when input #2 itself is a constant; or when input #1 is a constant and the instruction is commutative. Next, we check if it is possible to get the value of input #1 directly from memory. For that, input #1 must be an embeddable load and the output must be the corresponding embeddable store. If not, in case the instruction is commutative, we do the same tests for input #2. Finally, if none of the previous tests succeeded, we check if input #2 is an embeddable load, or if the instruction is commutative and input #1 is an embeddable load. In this case, the instruction can access the value of input #2 directly from memory.

In step 4, we deal with the following situations:

- Process a remaining embeddable load.
- Retrieve a spilled value.
- Move data between registers.
- Copy values that will be overwritten.

**Native Code Emission**

We extend Rotor's set of macros for code generation for IA-32, by adding macros that take advantage of the architecture's rich address modes. Namely, we add macros to use scale, index, and base when accessing arrays or data structures. We added a set of macros to emit logic and arithmetic instruction of type REG_MEM, MEM_REG, and MEM_IMM, which are not supported by FJIT.

Fig. 6 compares the code emitted by FJIT and AJIT, for the method *X::m* in Fig. 5. AJIT uses scaled, based and indexed access to the array. It also expands inline the tests for null reference and array bounds checks. Grey italic code is only executed in case of exception. Note how some instructions (marked with *) are not even emitted. Also note how FJIT uses almost exclusively register eax, while AJIT uses the register set more effectively.

```
public class X {
    private int[] v = new int[8];
    public int m(int a, int b) {
        return v[a * b + b];
    }
}
```

**Fig. 5.** Test code

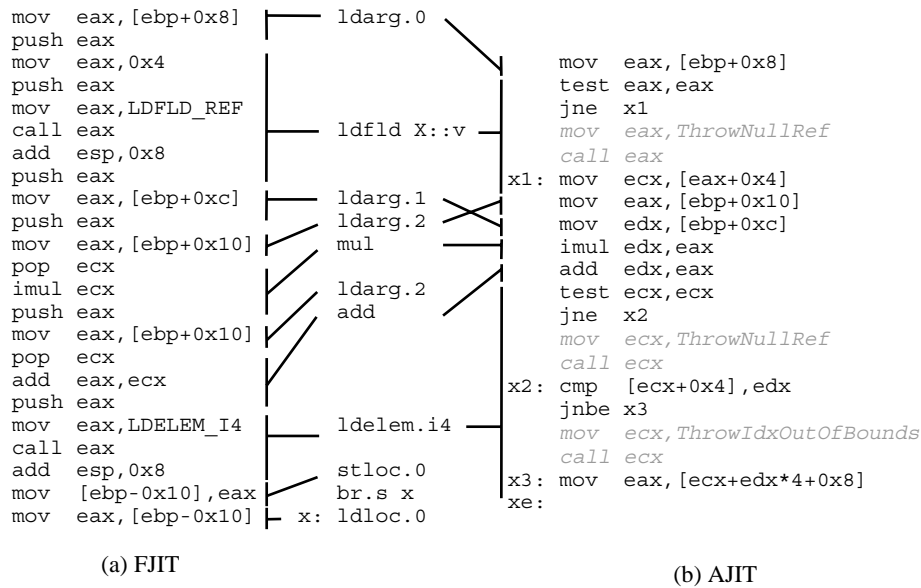

(a) FJIT  (b) AJIT

**Fig. 6.** Code generated by Rotor's JIT (a) and our JIT (b).

## Results

In this section, we compare AJIT with FJIT and, whenever possible, with CLR's JIT compiler. The programs used for this evaluation were developed considering AJIT's limitations. They implement two well-known algorithms: an iterative version of the sorting algorithm *QuickSort*; and an implementation of the RC4 [7] cipher algorithm.

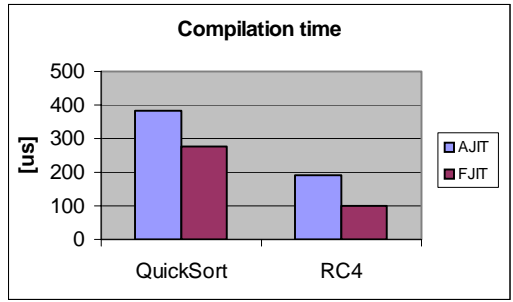

**Fig. 7.** Average compilation time

Fig. 7 shows that, as expected, AJIT is slower than FJIT. However, despite performing three major passes over the source code, AJIT does this in less than double the time spent by FJIT, which performs a single pass. On the other hand, FJIT does code verification during compilation, while AJIT skips it

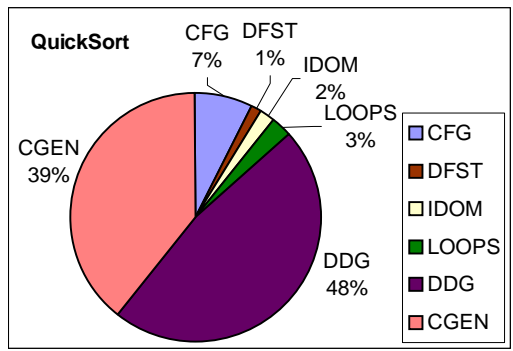

**Fig. 8.** Compilation time distribution for QuickSort

Fig. 8 provides a more detailed view of AJIT's compilation performance, illustrating the distribution of compilation time spent in each sub-phase for the *QuickSort* method. The results were similar for the RC4 implementation.

The code generation (CGEN) phase and data dependence graph (DDG) construction (which includes optimizations) are the two major contributors to total compilation time. Each of these two phases takes approximately 40% of total compilation time.

The control flow graph (CFG) construction, basic block depth-first numbering (DFST) and loop identification (IDOM and LOOPS) do not take more than 20% of total compilation time.

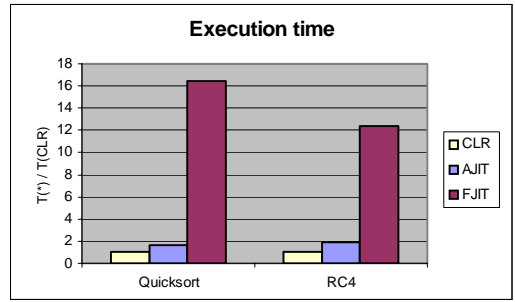

**Fig. 9.** Average execution time

Fig. 9 shows a comparison between the average execution time of the test methods when compiled with each of the JIT compilers (AJIT, FJIT, and CLR's). The execution time values are relative to the time obtained for methods compiled with CLR's JIT compiler, which has a reference value of 1. We verify that AJIT significantly improves the execution time of compiled methods, when compared to FJIT, and gets closer to CLR's compiler.

## Related Work

Besides Rotor [2] and the CLR [8], the two implementations of the CLI by Microsoft, we know about other three implementations: Novell/Ximian's Mono [9], DotGNU's Portable.NET [10], and Intel's Open Runtime Platform (ORP) [11] with StarJIT [12] plugged in. Both Mono and ORP use JIT compilers for the execution of CIL. Portable.NET, on the other hand, currently uses a two-level interpreter.

Mono has a single JIT compiler that uses an intermediate format based on SSA and allocates variables to registers with a linear scan algorithm. Mono's JIT optimizations take several passes over the code. StarJIT, a just-in-time compiler supporting both Java bytecodes and CIL, also uses an SSA-based intermediate representation. It implements lazy compilation, and uses a specialized register allocation algorithm. In lazy compilation, methods are first compiled with simple optimizations and later, if selected by the Profile Manager as frequently executed methods, they are recompiled with several optimization passes over the code.

In the Java world, the implementations of the Java Virtual Machine may be based on interpreting Java bytecodes [13] or on just-in-time compilation [14][15]. In order to achieve performance, the just-in-time compilers are usually simple and fast or use lazy compilation. In [3], Chen and Olukotun propose an alternative way to achieve high quality code with fast compilation times, by reducing the number of passes in the compiler to three, integrating the optimizations in these passes, and choosing the algorithms with a better trade-off in speed and quality. In our compiler, we explored a similar approach for CIL just-in-time compilation.

## Conclusions and Future Work

We have implemented the base of an optimizing just-in-time compiler for Rotor. This base includes the front-end, which translates CIL to an intermediate representation: a control flow graph, with the basic block structure of the program, and directed acyclic graphs, representing data dependencies. Additionally, the front-end determines the set of natural loops of the program. This information can be used for driving loop specific optimizations. We have also developed the code generator, which does register allocation simultaneously with native code emission.

Most compilation time is spent building the data dependence graph and generating native code. The construction of the data dependence graph includes a significant number of optimizations, which justifies the time it takes. On the other hand, with further work, we believe it is possible to reduce the time it takes to generate native code.

AJIT compiled methods execute significantly faster than those compiled with Rotor's original compiler, and are much closer to CLR compiled methods. Execution times are particularly encouraging as the developed compiler still lacks important optimizations like inlining, devirtualization, or alias analysis, and because all the implemented optimizations are only local to basic blocks. We also believe that, with further work, it is possible to reduce the time it takes to generate native code.

The work needed to complete AJIT can be classified in three areas: fully support CIL; total integration with Rotor; and additional optimizations. In order to fully support CIL, we need to deal with exceptions, value types, floating-point operations, and more. The most important step in order to complete the integration with Rotor is the interaction with the garbage collector in the following three points: write barrier; periodic polling; and tracing roots on the stack. Then, a lot of additional optimizations may be added, particularly the ones mentioned above and also loop specific ones.

## Acknowledgments

This work was developed in the context of the graduation thesis for the program in Computer Engineering at Instituto Superior de Engenharia de Lisboa. The authors would like to express their deep gratitude to professors Pedro Félix, Carlos Martins, and Luís Falcão for all the support and advising during the course of the project.

## References

[1] Standard ECMA-335: Common Language Infrastructure (CLI), 2nd ed., December 2002.
[2] David Stutz, Ted Neward, Geof Shilling, *Shared Source CLI Essentials*, O'Reilly, 2003.
[3] Michael Chen, Kunle Olukotun, "Targeting Dynamic Compilation for Embedded Environments", *Proceedings of the 2nd JavaTM Virtual Machine Research and Technology Symposium (JVM '02)*, San Francisco, United States of America, August 1-2, 2002.

[4] Keith D. Cooper, Timothy J. Harvey, Ken Kennedy, "A Simple, Fast Dominance Algorithm", Software Practice and Experience, 2001.
[5] Thomas Lengauer, Robert Endre Tarjan, "A fast algorithm for finding dominators in a flowgraph", *ACM Transactions on Programming Languages and Systems (TOPLAS)*, Vol. 1, Issue 1, July 1979, pp. 121-141.
[6] Andreas Krall, "Efficient JavaVM Just-in-Time Compilation", *IEEE Proceedings of PACT'98*, Paris, France, 12-18 October 1998.
[7] Bruce Schneier, *Applied Cryptography*, 2nd ed., John Wiley & Sons, 1996.
[8] Don Box, Chris Sells, *Essential .NET Volume 1 - The Common Language Runtime*, Addison Wesley, 2002.
[9] Mono Project, Novell/Ximian, http://www.mono-project.com
[10] Portable.NET, DotGNU Project, http://www.dotgnu.org/pnet.html
[11] Michal Cierniak et al., "The Open Runtime Platform: A Flexible High-Performance Managed Runtime Environment", *Intel® Technology Journal*, Vol. 07, Issue 01, February 19, 2003, pp. 5-18.
[12] Ali-Reza Adl-Tabatabai et al., "The StarJIT Compiler: A Dynamic Compiler for Managed Runtime Environments", *Intel® Technology Journal*, Vol. 07, Issue 01, February 19, 2003, pp. 19-31.
[13] JamVM, http://jamvm.sourceforge.net/
[14] "The Java Hotspot Performance Architecture", Sun Microsystems,
[15] http://java.sun.com/products/hotspot/whitepaper.html
[16] Kaffe, http://www.kaffe.org/